\documentclass{elsarticle}
\usepackage{ifpdf}

\usepackage{graphicx,natbib,amssymb,lineno}
\usepackage{epsfig}
\usepackage{epstopdf}
\ifpdf
\usepackage[%
  pdftitle={Instructions for use of the document class
    elsart},%
  pdfauthor={Simon Pepping},%
  pdfsubject={The preprint document class elsart},%
  pdfkeywords={instructions for use, elsart, document class},%
  pdfstartview=FitH,%
  bookmarks=true,%
  bookmarksopen=true,%
  breaklinks=true,%
  colorlinks=true,%
  linkcolor=blue,anchorcolor=blue,%
  citecolor=blue,filecolor=blue,%
  menucolor=blue,pagecolor=blue,%
  urlcolor=blue]{hyperref}
\else
\usepackage[%
  breaklinks=true,%
  colorlinks=true,%
  linkcolor=blue,anchorcolor=blue,%
  citecolor=blue,filecolor=blue,%
  menucolor=blue,pagecolor=blue,%
  urlcolor=blue]{hyperref}
\fi

\makeatletter
\def\elsartstyle{%
    \def\normalsize{\@setfontsize\normalsize\@xiipt{14.5}}
    \def\small{\@setfontsize\small\@xipt{13.6}}
    \let\footnotesize=\small
    \def\large{\@setfontsize\large\@xivpt{18}}
    \def\Large{\@setfontsize\Large\@xviipt{22}}
    \skip\@mpfootins = 18\p@ \@plus 2\p@
    \normalsize
}
\@ifundefined{square}{}{\let\Box\square}
\makeatother

\def\file#1{\texttt{#1}}

\begin{document}

\begin{frontmatter}
\title{Efficiency of Human Activity on Information Spreading on Twitter}

\author{A. J. Morales}
 %\address{Grupo de Sistemas Complejos and Departamento de F\'isica y Mec\'anica. Universidad Polit\'ecnica de Madrid. ETSI Agr\'onomos, 28040, Madrid, Spain}
\author{J. Borondo}
% \address{Grupo de Sistemas Complejos and Departamento de F\'isica y Mec\'anica. Universidad Polit\'ecnica de Madrid. ETSI Agr\'onomos, 28040, Madrid, Spain}
 \author{J. C. Losada}
% \address{Grupo de Sistemas Complejos and Departamento de F\'isica y Mec\'anica. Universidad Polit\'ecnica de Madrid. ETSI Agr\'onomos, 28040, Madrid, Spain}
 \author{R. M. Benito}
 \address{Grupo de Sistemas Complejos and Departamento de F\'isica y Mec\'anica. Universidad Polit\'ecnica de Madrid. ETSI Agr\'onomos, 28040, Madrid, Spain}

\ead{rosamaria.benito@upm.es}
%\ead[url]{authors.elsevier.com/locate/latex}

\begin{abstract}
Understanding the collective reaction to individual actions is key to effectively spread information in social media. In this work we define efficiency on Twitter, as the ratio between the emergent spreading process and the activity employed by the user. We characterize this property by means of a quantitative analysis of the structural and dynamical patterns emergent from human interactions, and show it to be universal across several Twitter conversations. We found that some influential users efficiently cause remarkable collective reactions by each message sent, while the majority of users must employ extremely larger efforts to reach similar effects. Next we propose a model that reproduces the retweet cascades occurring on Twitter to explain the emergent distribution of the user efficiency. The model shows that the dynamical patterns of the conversations are strongly conditioned by the topology of the underlying network. We conclude that the appearance of a small fraction of extremely efficient users results from the heterogeneity of the followers network and independently of the individual user behavior.
\end{abstract}

\end{frontmatter}

\section{Introduction}

In the recent years, our society has experienced the rise of new ways to communicate and relate among each other through digital devices. The increasingly affordability of technology, together with the solutions brought, have turn mobile and Internet devices as one of the fastest growing markets worldwide \cite{TelecomEconomist}. Specially in third world countries where the expanding projections of technological solutions double those found in the industrialized world \cite{BigDataThirdWorld}. Such technological revolution has given as a result, a massive amount of data provided by humans, as they interact with their digital devices on daily basis. The nowadays challenge is to turn these unstructured data into valuable information for policy makers to take better and more intelligent decisions  \cite{Lazer2009}.  

At the moment, traditional surveys have given important insights to our societal understanding. However, their cost in time and human efforts, makes it impossible for them to scale up and bring information of the structure of the social system behind their observation. Traditionally, the discovery of structural properties of social networks have been limited to the necessity of mapping a large amount of interactions between people. In this sense, online social networks, such as Twitter or Facebook, have become an ideal source of information to collect human-to-human interactions and unveil the social structures that people constitute, which opens an opportunity for researchers to characterize and model human behavior \cite{Lewis2008,Takhteyev2012}. These web applications are used on daily basis by people to post opinions, propagate news and exchange information. As a result, several commercial, political and social organizations are increasingly exploiting this communication tool to advertise products, organize campaigns and disseminate updates on their respective fields. 

Twitter, with over 200 million users, is the ideal tool to quickly propagate short text messages. It is an open debate that the data taken from Twitter are not necessarily representative samples of the outside world, as they are constrained to the population that participates in the online conversations \cite{Mislove2011,Gayo2012}. However, a social contextualization of the data, combined with a suitable computational and mathematical treatment, may provide important insights into how people behave. In fact, the activity performed by users on Twitter has brought information enough to understand a wide variety of phenomena, like the prediction of stock market variations \cite{Zeng2011}, the management of natural disasters \cite{Matsuo2011}, the understanding of epidemical diseases \cite{Culotta2010} and the characterization of electoral processes \cite{Borondo2012,Livne2011}. The deeply understanding of these social processes is crucial to design better strategies and get optimal outcomes from the network potential.

Recent studies have revealed that most of the information posted on Twitter is hardly propagated through the network, as 71\% of the messages do not travel any farther than the authors time-line \cite{Sysomos}. Among other factors, this spreading inertia has been attributed to the fact that the novelty of the posted information decays quite rapidly, which stretches the effective time to attract the collective attention \cite{Wang2011}, in addition to the fact that most of the people on Twitter behave passively \cite{Romero:2010tj}. However, in this context, there are people who do influence the rest of users and are able to get their messages spread through the network, in a wide variety of proportions.
% In fact, most of the exploiting strategies on Twitter are designed towards maximizing the number of the message readers and contacting the largest amount of possiMislove2011,Gayo2012}ble interested persons.

The keys to success when propagating information on Twitter have been reported to be a combination of several factors, such as the popularity of the source, the posting frequency, as well as the novelty and resonance of the message content \cite{Romero:2010tj}. In fact, the largest retweets cascades on Twitter, were found to be seeded by previously popular users, whose messages contained positive feelings \cite{Watts2011}. However, the {\color{black}efforts of} each user to gain influence and get their information spread on the network is a subject that has not yet been explained. In the sense, that although users may gain enough influence to transfer information on the network, this influence is not necessarily archived with the same efficiency, in terms of the amount of efforts that had to be employed for this matter.

In this work we address the question of which factors, like the individual behavior or the underlying substratum, determine the users efficiency to have their messages spread through the network. More specifically, we propose a measure to characterize the user efficiency to influence the emergence and growth of retweets cascades, by means of the relationship between the activity employed by the users and the emergent collective response to such activity, measured in terms of the number of retransmissions gained. On this basis, we propose a model to understand the emergence of the user efficiency distribution, based on independent cascades taking place on networks \cite{Goldenberg2001}, biasing the probability of retransmission among nodes, in order to decay as we move farther from the message source, as we see in the empirical data.

The results indicate that some regular users may gain a similar amount of retransmissions as the popular ones, but far less efficiently, as they must employ a much larger amount of activity. Furthermore, we have seen that the emergent distribution of users according to their efficiency, is strongly conditioned to the underlying network where information is being propagated. As a matter of fact, it actually represents a reflection of the dynamical rules behind the spreading process.

The paper is organized as follows. First, we introduce the system of our study in section \ref{SecSystem}, as well as the datasets that we have built and analyzed. Then {\color{black} in sections \ref{EmpiricalStudy}, \ref{eff} and \ref{Universality} }we focus on the empirical measurements that lead us to state the dynamical rules of the propagation process. After this, in section \ref{Model} we propose a simple model to verify the dynamical processes reported. Finally, we discuss the effects of the underlying topology and initial user activity behavior in the emergent dynamical patterns, which we found to be universal on Twitter conversations.

\section{System}
\label{SecSystem}

The system under study is based on human activity taking place around specific topics of conversation on Twitter. In this section we give some background on the user interaction mechanisms provided by Twitter, as well as describe the datasets that we have built and analyzed.

\subsection{Twitter Background}

Twitter is a microblogging service where people are able to post and exchange text messages limited by $140$ characters either from personal computers or mobile devices. There are several mechanism for users to interact on Twitter. The first of these is the ability to follow and be followed by other persons. This is a passive mechanism that allows users to receive all the messages posted by those who follow, as well as to deliver their own messages to their own followers. In this sense, it establishes the Twitter followers network, where the users are connected among each other, through links that determine the explicit ways where messages are delivered. Previous studies have reported complex properties in this network \cite{Kwak2010}, like degree distribution with power law behavior, small mean distance between nodes and modular structure. However, it has been observed that individuals do not actively interact with all of the declared contacts, but only with a small fraction of them \cite{Huberman2009}. Among these active mechanisms to interact, the {\it retweet} (or retransmission) is the most popular one to propagate the received messages throughout the network. By retweeting a message, users deliver specific information to their own followers, at the same time that endorse ideas and gain visibility in the network \cite{Lotan2010}. The study of the retweets cascades has served to characterize user profiles \cite{Despotovic2010}, measure influence \cite{Cha2010} and propose spreading models \cite{Xiong2012}. At last, all messages on Twitter, may be identified using keywords called {\it hashtag}. This mechanism organize conversations and individuals use it to exchange ideas on specific subjects. Recently, the statistical analysis of the hashtags usage has let prediction on social relations \cite{RomeroHT} and collective attention \cite{LehmannHT}.

\subsection{Datasets}
\label{SecSystemDataset}

 {\color{black}Using the Twitter Search API version 1.0  \footnote[1]{https://dev.twitter.com/docs/using-search}, we have built several datasets from public access messages. This API provides data from a temporal index of recent tweets, posted within a lapse of a week from the time the query is made. The limitations of this API are not specified as a relative volume of messages, nor a fixed number of queries, but instead a combination of the queries' complexity and frequency. The datasets were built querying for messages with specific keywords related to topics of conversation that captured a significant part of the collective attention. Their sizes vary from $10^4$ to more than $10^6$ messages or participants, as may be seen in Table \ref{HT}.}

First, we considered an online Venezuelan political protest as a case study. This event took place exclusively on Twitter on Dec. 16th, 2010. Two days before the protest, the convoker asked his followers to post messages identified with the hashtag \#SOSInternetVE, who responded massively and the conversation propagated becoming trending topic. We collected up to 421,602 messages, identified with the protest hashtag, which were posted by 77,706 users, between Dec. 14-19, 2010 (two days before and after the protest). In our previous work \cite{Morales2012}, we found that some influential users acted as information producers, providing messages that are received by the passive large majority of information consumers. Besides, we found that users are organized in a community structure around hubs of different nature, like politicians, humorists or mass media accounts.

Second, in order to generalize results, other datasets were also built around other conversation topics of different nature such as sports, news, protests and political campaigns. The first of these datasets is related to a political scandal that took place on the Spanish parliament on 2012 due to some unappropriated comments from a congresswoman that echoed loudly on the social networks. This dataset was built by downloading the hashtag {\it \#Andreafabra}, which corresponds to this person\rq{}s name, from July 12th, 2012, to July 23th, 2012. The second dataset concerns a conversation about a Venezuelan baseball team. It was built by downloading the messages that contained the team\rq{}s name {\it leones} during a 3 weeks period from Dec. 22th, 2010, to Jan. 12th, 2011. Moreover, we have built another dataset concerning the 2011 Arab Spring, by downloading the messages that contained the keyword (and hashtag) {\it Egypt} during a 5 week period, from Jan. 12th, 2011, to Feb. 17th, 2011. During this period the former Egyptian president was overthrown by the social revolts. Besides, two datasets concerning the American 2012 elections were built by respectively gathering all the messages that contained the word {\it Gingrich} during a week period from Feb. 29th, 2012, to Mar. 3rd, 2012, as well as the word {\it Obama} during the first televised debate from Oct. 3th, 2012, to Oct. 5th, 2012. Finally, the last of these datasets is related to the 2011 Spanish electoral process. It has been built with all the messages that contained the keyword (and hashtag) {\it 20N}, which was used by all parties in reference to the election day on Nov. 20th, 2011. This dataset comprehends the period from Oct. 29th, 2011, to Nov. 27th, 2011. In our previous work of this electoral process \cite{Borondo2012}, we characterized the user and politicians interactions and found that the mass media accounts widely dominated the attention received through the retweets mechanism, while politicians ruled the mentions scenario.

\begin{table}
\begin{center}
\caption{Properties of the studied datasets and their resulting user efficiency distribution properties.}
\label{HT}
\begin{tabular}[c]{|c|c|c|c|c|}
\hline Keyword & Messages & Users & $\mu_{\eta}$ & $\sigma_{\eta}$ \\
%\hline La\_upm & $2,840$ & $1,746$ &$-0.23$ & $1.07$\\
%\hline Eurocrisis & $5,846$ & $6,301$ & $0.15$ & $1.18$\\
\hline Andreafabra & $35,835$ & $23,498$ & $0.15$ &$1.05$\\
\hline Gingrich & $93,063$ & $43,061$ & $-0.08$ & $1.13$\\
\hline Leones & $142,808$ & $46,608$ & $-0.08$ & $1.09$\\
\hline 20N & $389,988$ & $123,710$ & $-0.49$ & $1.08$\\
\hline SOSInternetVE & $421,602$ & $77,706 $ & $-0.79$ & $1.21$\\
\hline Obama & $6,818,782$ & $2,265,799$ & $0.14$ & $1.15$\\
\hline Egypt & $7,433,542$ & $1,180,715$ & $-0.80$ & $1.33$\\
\hline
\end{tabular}
\end{center}
\end{table}

\section{Characterizing the Spreading Behavior}
\label{EmpiricalStudy}

In this section we present the overall behavioral patterns of the conversation \#SOSInternetVE. We analyze the user activity, as well as the underlying social network and the emergent retweet network.

\begin{figure}
\begin{center}
 % replace aims_logo.eps by your figure file name
 \includegraphics[width=4in]{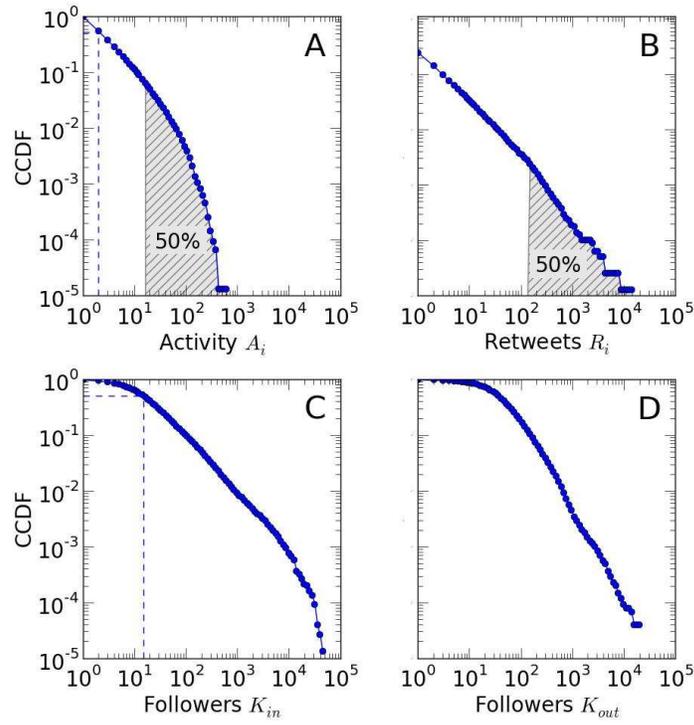}\\
 \caption{Complementary cumulative density function (CCDF) of (A) user activity, measured as  the number of messages sent by user; (B) in strength of the retweet network; (C) in degree of the followers network and (D) out degree of the followers network. The dashed line marks where about half of the users are located in the distribution and the gray regions determine the area that covers half of the samples. The distributions correspond to the \#SOSInternetVE dataset. }\label{System}
 \end{center}
\end{figure} 

%In this section we will discuss the user behavioral patterns concerning their activity and received attention, in terms of audience (followers) and propagators (retweeting behavior). The analyses are performed on the Venezuelan protest dataset previously described in section \ref{SecSystemDataset}.

\subsection{Activity Behavior}
\label{ActivityB}
The user activity $A_i$ is considered as the sum of the original and retransmitted messages, sent by each participant $i$.  {\color{black} Its complementary cumulative density function (CCDF) presents a broad distribution, as can be seen in Fig. \ref{System}A, which means that users participated quite heterogeneously in the conversation. }This distribution indicates that up to 53.3\% of the participants posted at most two messages each (dashed line in Fig. \ref{System}A), which represents less than 10\% of the total messages posted, while the remaining 90\% of the messages were sent by almost the other half of the population (46.7\%), who posted more than two messages by person. The conversation stream was actually fed from a small group of the most active users (6\% of the participants), who individually posted from 16 to around 630 messages, and whose activity represent half of the overall amount of messages (shadow region in Fig. \ref{System}A). Previous studies on Twitter \cite{Sysomos}, attribute 75\% of the overall messages to 5\% of the entire population, which indicates that an unusual high amount of users participated in this protest.

\subsection{Followers Network}

In the same manner that users post messages quite differently among them, these messages have also different relevance in the conversation development. On Twitter, not all the users account the same level of visibility in the message stream, because the number of recipients, and possible readers, strongly depends on the source\rq{}s in degree on the {\it followers network} (see Sec. \ref{SecSystem}). This social substratum may be analyzed by the construction of a graph with the protest participants, linking the users according to {\it who follows who}. The resulting is a directed and non weighted network compound by 77,706 nodes and 5,761,331 links, displaying the structure through which information is delivered and might be spread. The edge direction goes from the follower to the message source, thus information flows in the opposite sense of the edges and therefore the attention received can be measured by means of the in degree $k_{in}$. As it can be seen in Figs. \ref{System}C-D, the in and out degree distributions of the followers network present power law behavior above three orders of magnitude, which is a property of scale-free networks \cite{Newman2006}. This indicates that while 51.7\% of the population is followed by less than 15 users (dashed line in Fig. \ref{System}C), there exist a very few accounts, like the protest convoker, who are followed by over 40.000 users, which correspond to more than half of all the participants. These popular accounts are mainly related to mainstream, celebrities, politicians or popular bloggers, and whose messages are widely received among the protest participants.

\begin{center}
\begin{figure*}

 % replace aims_logo.eps by your figure file name
 \includegraphics[width=4.7in]{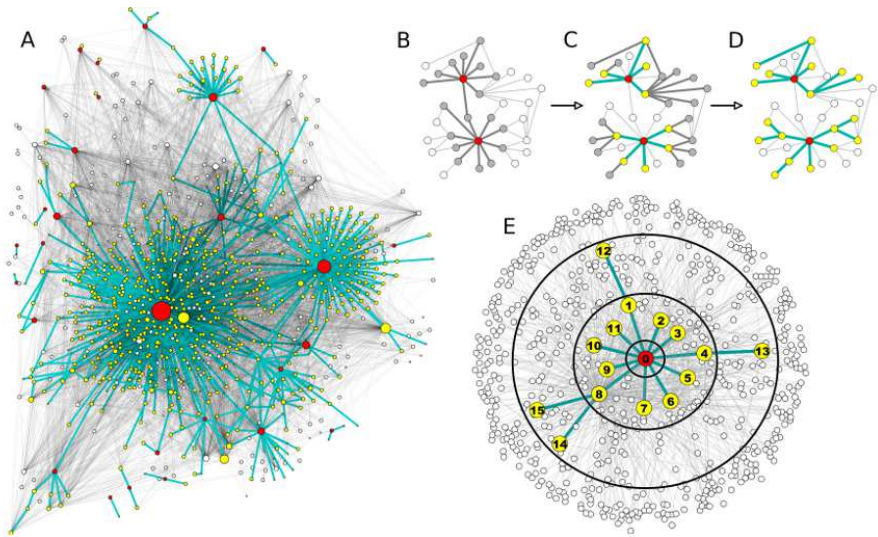}\\
 \caption{Visualization of the retweet network emergent from the message propagation on the followers network. (A) Subgraph of the retweet network (green) superimposed to the corresponding followers network (black), from the \#SOSInternetVE dataset. In the figure a subset of 1000 random nodes (yellow and red) are presented. The node size is proportional to the respective in degree on the followers network. (B, C and D) Example of the formation of the retweet network from independent retweet cascades on an artificial followers network. (B) shows when two users (red nodes) post independent messages which are received by their followers (gray). (C) shows when some users retweeted the message (yellow) and this message arrives to their followers (gray). (D) shows the final shape of the cascades on the network, compound only by the activated nodes (red and yellow) connected by the green links. The white nodes and gray links represent the rest of the substratum (followers network) who did not activate. (E) shows the schema of a single cascade. The black circles determine the cascade layers.}\label{Figure2}

\end{figure*}
 \end{center}

In order to unveil how these heterogeneous users interacted with each other, we calculated the assortativity by degree coefficient \cite{Newman2003} for this followers network. The network resulted to be disassortative ($r = -0.10$), which reveals the asymmetric configuration, where the hubs that concentrate much of the incoming links, are often targeted by regular users, who do not receive much of the collective attention. Although social networks have been reported to be assortative \cite{Newman2003}, this pattern changes in the online world, where disassortativity is usually found \cite{Hu2009}. This is due to the new mechanisms that allow regular people to interact and communicate with popular accounts, like following them in the case of Twitter.

\subsection{Retweets Network}
\label{ActivityR}

The heterogeneous behavior of the followers network, gives place to a high level of disparity in the reception of the messages and consequently in the information spreading process. To further understand  it, we analyzed the {\it retweet network} that emerged from the mentioned conversation. In this network nodes represent users, and edges are created according to {\it who retransmits whose messages}. The edges are directed and weighted according to the number of times users retweeted each other, plus the number of subsequent propagators that retweeted the same message. This network can also be seen as the aggregation of independent {\it retweet cascades}, that respectively occur when a single message is retransmitted by any user to its followers, allowing them and their own followers, to do the same. An example of the resulting structure is shown in Fig. \ref{Figure2}A, where a subset of the retweet network (green edges) has been plotted, superimposed to the respective subgraph of the followers network (gray edges). The red nodes represent those who posted an original message and the yellow nodes represent the message propagators (those who retweet). It can be noticed that the retweet network represents a subset of the followers graph where messages are actually being propagated. This graph evidences that people are more selective to actively interact with their declared contacts than just receiving updates from them \cite{Huberman2009}.

In order to explain the dynamical process behind these cascades, an scheme of the evolution of two cascades on an artificial followers network is sketched from panels B to D in Fig. \ref{Figure2}. In panel B two independent messages are respectively posted by the red nodes and received by their followers (gray nodes). Some of these followers retransmitted the messages (yellow nodes), through the green edges, and others did not (white nodes), as shown in panel C. Accordingly, in panel D some of the followers of followers retransmitted the message (also yellow nodes), and the final shape of the cascades may be appreciated. To summarize it schematically, a single retweet cascade from the dataset is presented in Fig. \ref{Figure2}E. The white nodes do not belong to the cascade, as we only consider those who actively participated in the retransmission process. Using this schema some of the main cascade properties will be explained in the remaining section, such as the amount of retransmissions gained by user, as well as the cascade size, depth and rate of retransmission.

The first property we analyzed is the number of retweets gained by user, $R_i$, which may also be considered as the node $i$ in strength of the retweet network. This quantity may increase either from cascades originally seeded by $i$, as well as cascades where $i$ acted as a propagator. For example, for the cascade shown in Fig. \ref{Figure2}E, $R_i$ would take the following values: $R_0 =15$, which is the total number of users who retweeted the message originally posted by the node $0$, either directly (nodes $1$ to $11$) or indirectly (nodes $12$ to $15$). Accordingly, $R_8 = 2$, since the node 8 has been retweeted by nodes 15 and 14; $R_1 = R_4 = 1$, since node 1 and 4 have been retweeted by node $12$ and $13$ respectively; and finally $R_2 = R_3 = R_5 = R_6 = R_7= R_9= R_{10}= R_{11} = 0$, as no one retweeted them. 

In Fig. \ref{System}B, we present the results of $R_i$ for the considered conversation. It can be noticed that $R_i$ is distributed following a power law behavior, where only 25\% of the overall users got retweeted at least once. This means that those messages from the remaining 75\% of users had no effect on the growth of the retweet network. In fact, this network is widely dominated by 0.4\% of the participants, who concentrated half of the sum of the users $R_i$ (shadow region in Fig. \ref{System}B). After identifying who represent these influential accounts, we found them to be compound by popular users, who often appear in the traditional media and catalyze the diffusion of opinions behavior, as well as concentrate most of the collective attention.

\begin{figure}
\begin{center}
 % replace aims_logo.eps by your figure file name
 \includegraphics[width=4in]{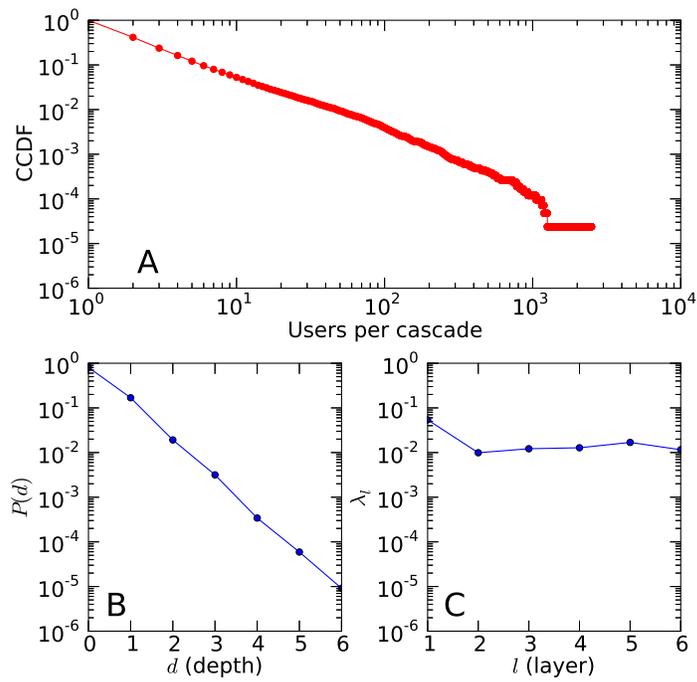}\\
 \caption{Retweets cascades statistical properties. (A) Complementary cumulative density function of the number of users per cascade, (B) Cascade depth distribution $P(d)$ and (C) Retransmission rate by layer $\lambda_l$ in terms of retweets over followers. The data correspond to the \#SOSInternetVE dataset.}
\label{CaracterizacionCadena}
 \end{center}
\end{figure}

Another property analyzed is the cascade size, which is defined as the total amount of nodes that have been activated in the context of a given cascade. In the example shown in Fig. \ref{Figure2}E the resulting cascade size would be 16, as we have 1 author (node 0) plus 15 propagators (nodes 1 to 15). In the studied conversation, this property is distributed following a power law behavior, as presented in Fig. \ref{CaracterizacionCadena}A. This indicates that most of the cascades are extremely small, as more than half of them (60\%) are compound at most by 2 persons besides the author, and just a small fraction are large, since around 5\% of them have more than 10 users, and 0.03\% present more than 100 participants. 

In order to understand the cascades structure, we have divided them by layers, as shown with the black circles in Fig. \ref{Figure2}E. The cascade layer indicates the number of hops from a propagator node to the source node, through the cascade links. The users correspondent to the layer $l=n$ represent those who retransmitted the message coming from a user of the previous layer $l=n-1$. In Fig. \ref{Figure2}E, the message author (red node) stands alone in the layer $l=0$, while in the consequent layers, we find those nodes who retweeted the message, like the nodes 1 to 11 in layer $l=1$, and the nodes 12 to 15 in layer $l=2$.

The cascade depth $d$ corresponds to the farthest layer from the message source, in which a node has been activated. In the example shown in Fig. \ref{Figure2}E, it would take the value of $d=2$. In the analyzed conversation, the probability of a cascade to have a certain depth, $P(d)$, is presented in Fig. \ref{CaracterizacionCadena}B. Those cascades of depth $d=0$, represent original messages that were not retweeted by anyone, which comprehends close to 80\% of them. In this sense, only 17\% of the cascades just have one layer of retransmission ($d=1$), and this quantity decreases exponentially as we move farther from the message's source, reaching a maximum depth of $d=6$ layers with a very low likelihood ($\sim 10^{-5}$). This indicates that the retweets cascades found in this conversation are quite shallow, which might result counterintuitive, as we would expect retransmissions to increase directly to the message's visibility, which should increase with each retransmission. However, shallow cascades have been detected on Twitter in works of influence dynamics \cite{Watts2011}  and prediction of urls propagation \cite{Despotovic2010}, as cases of different media, like the flow of emails inside a corporation \cite{Barabasi2011}. It has been shown that information tends to loose its capacity to attract attention when we move farther from the author\rq{}s social surroundings, and hence the probability of a cascade to grow is inversely dependent on the distance from the source node \cite{Wu2004}.

Finally, the rate of retransmission at each layer, $\lambda_l$, is estimated by averaging the ratio between the number of users who retransmitted a message normalized by the number of individuals who received it at each layer, taking into account the followers network information. The results are shown in Fig. \ref{CaracterizacionCadena}C, and it shows that $\lambda_l \sim 0.01$ for $l > 1$, while in the first layer the average retransmission ratio reached up to 5\% ($\lambda_l \sim 0.05$) of the exposed users.

\begin{figure}
\begin{center}
 % replace aims_logo.eps by your figure file name
 \includegraphics[width=4in]{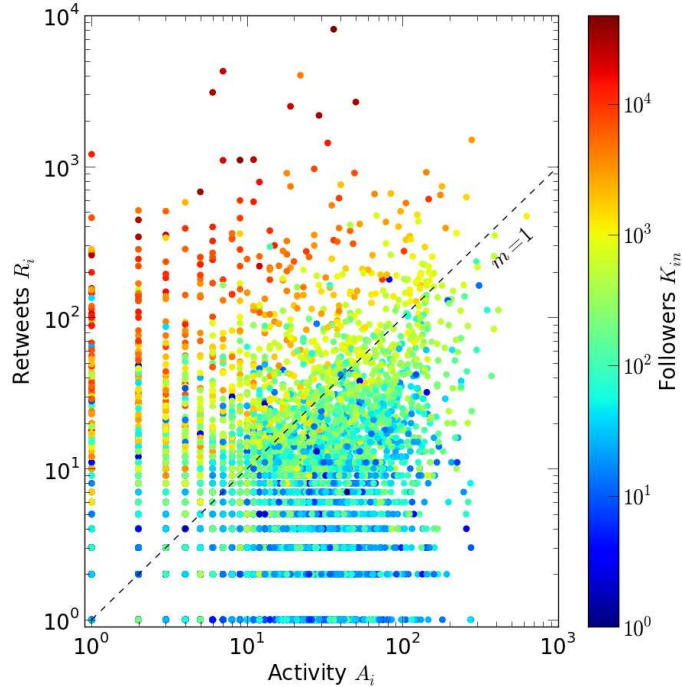}\\
 \caption{Scatter plot of the retransmissions gained by user versus its activity and colored by its number of followers. Dots represent users. Data correspond to the \#SOSInternetVE dataset.}\label{Figure3}
 \end{center}
\end{figure} 

\begin{table}
\begin{center}
\caption{Pearson correlation (r) by user of the number of followers (F), retweets (R) and activity (A), from the \#SOSInternetVE dataset.}
 \label{NetCorr}
\begin{tabular}{l l l l}
\hline\noalign{\smallskip}
Topic & $r_{F,A}$&$r_{F,R}$ & $r_{R,A}$\\
\noalign{\smallskip}
\hline
\noalign{\smallskip}
SOSInternetVE & $0.07$ & $0.57$ & $0.17$\\
\hline
\end{tabular}
\end{center}
\end{table}

\begin{figure}
\begin{center}
 % replace aims_logo.eps by your figure file name
 \includegraphics[width=4in]{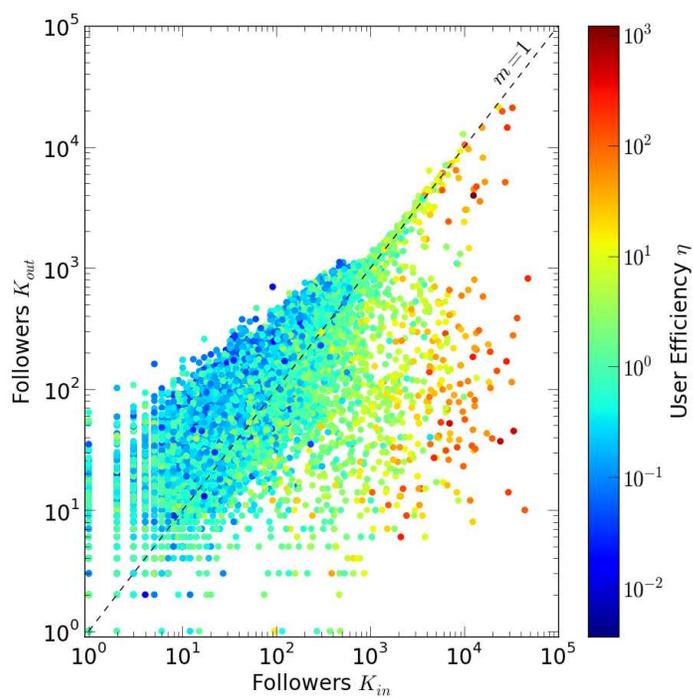}\\
 \caption{Scatter plot of the user in degree vs out degree in the followers network, colored by the respective user efficiency. Dots represent users. Data correspond to the \#SOSInternetVE dataset.}\label{Figure3B}
 \end{center}
\end{figure} 

\section{Efficiency of Human Activity}
\label{eff}

At this point it has been shown a significant heterogeneity in the users behavioral patterns, in terms of the activity distribution (number of messages posted) and the attention received (number of followers and retweets gained). However, the way these measures are correlated, and their relation to the user efficiency to spread information remains unanswered.

In Table \ref{NetCorr}, the Pearson coefficient between the users number of followers $F$ (measured as the $k_{in}$ in the followers network), retweets gained by user $R$ and activity $A$, are presented. It can be noticed that there is no correlation between the number of followers and activity employed ($r_{F,A} = 0.07$), which means that the amount of messages posted is independent of the user position in the followers network. However, there is a strong correlation between the number of followers and the retransmissions gained ($r_{F,R} = 0.57$), which means that the most retransmitted users tend to be the most followed ones as well. Besides, there is a positive correlation between the number of retransmissions and activity employed ($r_{A,R} = 0.17$), which indicates that the chances of being retransmitted increase with every message posted for all users. 

In Fig. \ref{Figure3}, we present a scatter plot of the retweets gained by user as a function of its activity and colored by the user $k_{in}$ in the followers network. It can be clearly noticed that the most retransmitted users are also the most followed ones (red dots), independently of their activity. However, some less followed users (green or yellow dots) may also gain a significant amount of retransmissions, but by means of a considerable increase in their own activity. These users are located around the straight line of slope 1, and their retransmissions gained are proportional to their activity. Finally, some not so followed users (blue dots in Fig. \ref{Figure3} below the dashed line), who are vast majority of the population, needed to post an enormous amount of messages to gain, if any, a few retransmission at most.

The fact that not all the participants must employ the same amount of effort, to accomplish the same level of retransmissions, implies that users have an individual efficiency to get their messages spread by others. This {\it user efficiency}, $\eta$, may be understood as the ratio between the collective response to the individual efforts. It is a metric of influence in the network, quantified as the amount of retransmissions gained by user with each message posted, defined according to the following expression: 

\begin{equation}
 \eta_i = \frac{R_i}{A_i}
\label{Nu}
\end{equation}

where $R_i$ is the number of retweets gained by user $i$, and $A_i$ is the amount of messages posted or retweeted by the user $i$. Those users whose $\eta > 1$ get more retweets than the number of messages posted and therefore are more efficient to spread their information in the network and consequently gain more influence, in comparison to those users whose $\eta < 1$, that employed larger efforts to obtain similar outcomes.
%The $\eta_i$ provides information of the expected aggregated effects, caused in the network, by the user\rq{}s individual actions.

\begin{figure}
\begin{center}
 % replace aims_logo.eps by your figure file name
 \includegraphics[width=4in]{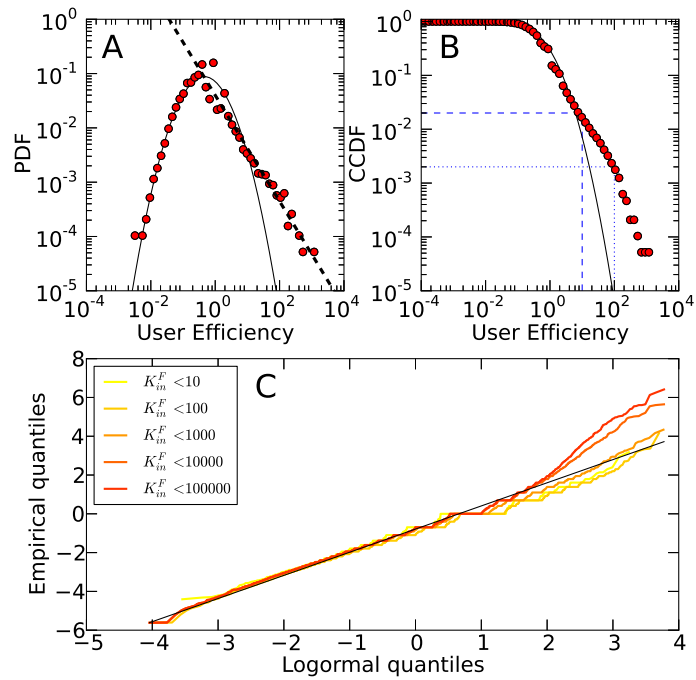}\\
 \caption{User efficiency probability density function (A) and complementary cumulative density function (B). The red dots correspond to the empirical results, the black solid line represents the lognormal fit and the black dashed line represents a power law fit.  Quantile-Quantile plot (C) of the user efficiency distribution, filtered by the in degree in the followers network $K_{in}^F$. The distributions correspond to the \#SOSInternetVE dataset.}\label{campanalog}
 \end{center}
\end{figure}

In Fig. \ref{Figure3B}, we present a scatter plot of the users degree in the followers network, $k_{in}$ and $k_{out}$, colored by their efficiency $\eta$. It may be noticed, that the users who present an efficiency $\eta > 1$ (green, yellow, orange and red dots) are mostly located below the dashed line of slope one, which means that their audiences ($k_{in}$) are larger than their sources of information ($k_{out}$), which implies a certain level of popularity in the network. Specially, those whose  $\eta >> 1$ (orange and red dots), who may be followed by more than $10^4$ users, but they only follow less than $10$ users. Meanwhile, the users who present a low efficiency (blue dots), tend to receive messages from much more sources than the size of their audiences ($k_{out} > k_{in}$), and also have a smaller amount of followers. This means that these users hear more information from the network, than what they are actually listened.

However, the mean efficiency value seems to be close to 1 ($R_i \sim A_i$), as shown in the user efficiency $\eta$ distribution presented in Fig. \ref{campanalog}A, which means that in average most of the users who got retweeted, gained as many retransmissions as the amount of messages posted. Besides, the users whose $\eta >> 1$, represent a minority part of the population, as clearly shown in the $\eta$ complementary cumulative distribution in Fig. \ref{campanalog}B. It can be noticed that less than 2\% of the retweeted population gained more than 10 retransmissions by message sent (dashed line in Fig. \ref{campanalog}B), 0.2\% gained over 100 retransmissions by message sent (dotted line in Fig. \ref{campanalog}B) and just one user gained over 1000 retransmissions with a single post.

{\color{black}In order to further understand the $\eta$ distribution, we have superimposed in Fig. \ref{campanalog}A-B the correspondent lognormal curve, with the mean and variance taken from the empirical observations (see Table \ref{HT}). It is known that lognormal distributions arise from multiplicative growing processes, like branching processes, as they may be explained by the central limit theorem, in the logarithmic scale \cite{Mitzenmacher2004}. An example of these processes are found in viral marketing campaigns \cite{Moro2011,Moro2011b}, where the number of leaves grow multiplicative as the branches split like the cascades shown in section \ref{ActivityR}. It can be noticed that the initial part of the distribution fits quite well the lognormal curve, but right after its maximum the distribution changes the scaling behavior, apparently to a power law, which we have also superimposed in Fig. \ref{campanalog}A with a dashed line. This means that there is a higher concentration of users who gain a larger amount of retransmissions by message posted, than what is expected for a lognormal distribution. These highly efficient users correspond to the hubs of the followers network as can be appreciated in Fig. \ref{campanalog}C, where we have plotted the Quantile-Quantile plot of the $\eta$ distribution in comparison to the lognormal distribution, filtered by the number of followers. If $\eta$ would follow a lognormal distribution, all the points would appear in a straight line, which actually happens for the users who present less than 1000 followers. But, as we consider the most followed users, the curve begins to change its behavior, suggesting that the underlying network topology is responsible for such deviation. This point would be further analyzed in section \ref{Results}.}

In summary, we have seen two kind of users who may gain a significant amount of retransmissions. One of them, are the highly connected users in the followers network, which have no need to follow other people, and with a high efficiency, gain a much larger amount of retweets than their own messages. While, there are other not so well connected users, who may also gain a lot of retweets, but in a less efficient way, since they need to post much more messages than the highly efficient ones.

\section{Universality}
\label{Universality}
\begin{figure}
\begin{center}
 % replace aims_logo.eps by your figure file name
 \includegraphics[width=4in]{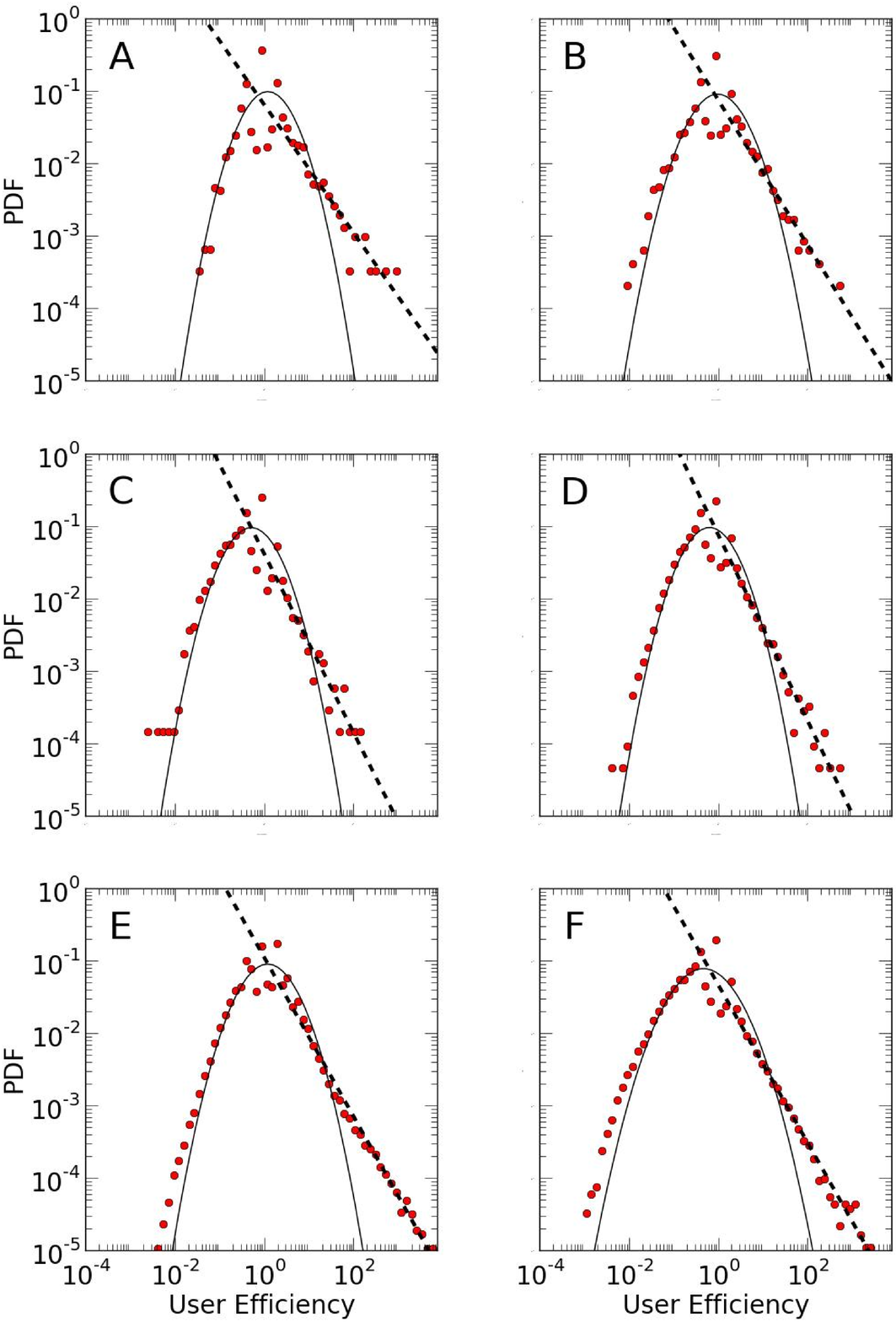}\\
 \caption{Probability density function of the user efficiency on several Twitter conversations, ordered increasingly according to the number of messages (A-F): (A) Andreafabra, (B) Gringich, (C) Leones, (D) 20N, (E) Obama, and (F) Egypt. The properties of these conversations may be found in Table \ref{HT}. The black solid line represents the lognormal fit, the black dashed line represents a power law fit and the red dots correspond to the measured distributions.}\label{RiAall}
 \end{center}
\end{figure} 

In order to identify whether this distribution is constrained to the present case study or rather represents a consequence of an universal feature of the interaction mechanism, we have calculated the user efficiency ($\eta$) for other conversations on Twitter. Specifically, we performed the analysis over six different datasets described in section \ref{SecSystem} and whose features may be found in Table \ref{HT}. All of them belong to different contexts and their sizes include several order of magnitude in terms of the number of posted messages and participant users. The results of the emergent $\eta$ distributions from these datasets are presented in Fig. \ref{RiAall}, plotted in ascendant order according to their size (from A to F). It can be noticed that the lognormal distribution emerges, even when the smallest datasets are considered (Fig. \ref{RiAall}A-B). However, as the size of the dataset increases, the effects of the presence of highly efficient users is more evident in the distributions, which present a very similar shape as the one found for the \#SOSinternetVE conversation (Fig. \ref{campanalog}A).

Given the fact that the size of the datasets cover from four to six orders of magnitude and correspond to topics of different nature, it is remarkable that the resulting distributions present a very similar shape. This ubiquity of the resulting patterns, strongly suggests the existence of an universal behavior in the relation between the individual efforts, managed by the user, and the collective reaction to such efforts, which is an emergent property of the underlying network. So we open the following question: what factors cause the emergence of such distribution? In the next section we will propose a model to explain the emergence of the observed distribution.

%This kind of distributions have been observed in related phenomena such as the global growth of conversations on Twitter \cite{Wang2011}, and the number of cascades generated by posted content, on another online social network \cite{Lerman2011}. 
%These processes have served to model several information cascades on electronic media, such as viral marketing campaigns \cite{Moro2011}. 

\section{Model}
\label{Model}

In order to model the propagation of retweets that took place on the \#SOSInternetVE conversation, we propose a spreading mechanism based on independent cascades \cite{Goldenberg2001} taking place on the followers network. In this model, nodes are activated in analogy to having posted a message, allowing their neighbors to also activate, like having retransmitted the received message, following the cascade schema shown in Fig. \ref{Figure2}. Each message may trigger an independent cascade regardlessly of the author\rq{}s previous activations. Besides, nodes may belong and participate in several cascades at the same time.

In the context of a given cascade, when a node $i$ has been activated, it has a single chance to activate each of its neighbors (followers), $j$, located at $l$ layers away from the message source. Thus the spreading probability depends on such distance $l$. In the sense that, the probability of a node $j$ to retransmit a message at $l$ layers away from the source, is given according to the probability of the cascade to grow vertically and have a depth of at least $l$ layers, $P(d \ge l)$, and the probability to grow inside the layer $l$, given by $\lambda_l$.

The user activity $A_i$ is given as the result of all the messages posted by $i$: as a source in layer $l=0$ ($A_{i,0}$) plus all the retweets made by $i$  at $l$ steps farther from the message source ($A_{i,l}| l>0$), in the following way:

\begin{equation}
A_i=A_{i,0}+\sum_{l=1}^{d_{max}}{A_{i,l}}
\end{equation}

where $d_{max}$ is the maximum cascade depth allowed. On one hand, $A_{i,0}$ is an independent random variable with density distribution $P(A_{0})$, and represents the initial conditions for the spreading process. On the other hand, $A_{i,l}| l>0$ is not independent and it rather represents a consequence of the propagation of other nodes\rq{} activity. Among other factors, this quantity depends on the amount of messages received by $i$, which is proportional to the amount of people who $i$ follows on the underlying followers network ($k_{i,out}$).

From this perspective, we define the retransmissions gained by user $i$ in the following way:

\begin{equation}
\label{eqRi}
 R_i=\sum_{l=0}^{d_{max}-1}{R_{i,l}}
\end{equation}

where $R_{i,l}$ represents the retweets gained by the node $i$ due to its given activations at the layer $l$ in all the cascades. This means that a node $i$ may gain retransmissions either from the messages originally posted by it ($R_{i,0}$), as well as from messages retweeted by $i$ at $l$ layers away from the source ($R_{i,l}$). On this basis, the value of $R_{i,l}$ depends on the number $i$\rq{}s followers, as well as the followers of followers, and so on, until reaching the maximum depth considered for a possible node activation, given by $d_{max}$. Hence the sum upper limit in eq. \ref{eqRi} is one layer before this value.
%The resulting user efficiency $\eta_i$ would be:

%\begin{equation}
%\label{eta2}
%\eta_i =\frac { \sum_{l=0}^{d_{max}-1} {R_{i,l} } }{\sum_{l=0}^{d_{max}} {A_{i,l}}}
%\end{equation}

In order to simulate the model, we must define the underlying network where the propagation process would take place, as well as the initial user activity distribution $P(A_0)$. Then the messages are spread taking into account the probability of a cascade to reach $l$ layers $P(d \ge l)$ and the retransmission rate in a given layer $\lambda_l$. Finally after all the initial activations are performed and the triggered cascades extinct, we calculate the efficiency $\eta$ for each user according to eq. \ref{Nu}, as well as the correspondent density distribution.

\subsection{Results}
\label{Results}
\begin{figure}
\begin{center}
 % replace aims_logo.eps by your figure file name
 \includegraphics[width=4in]{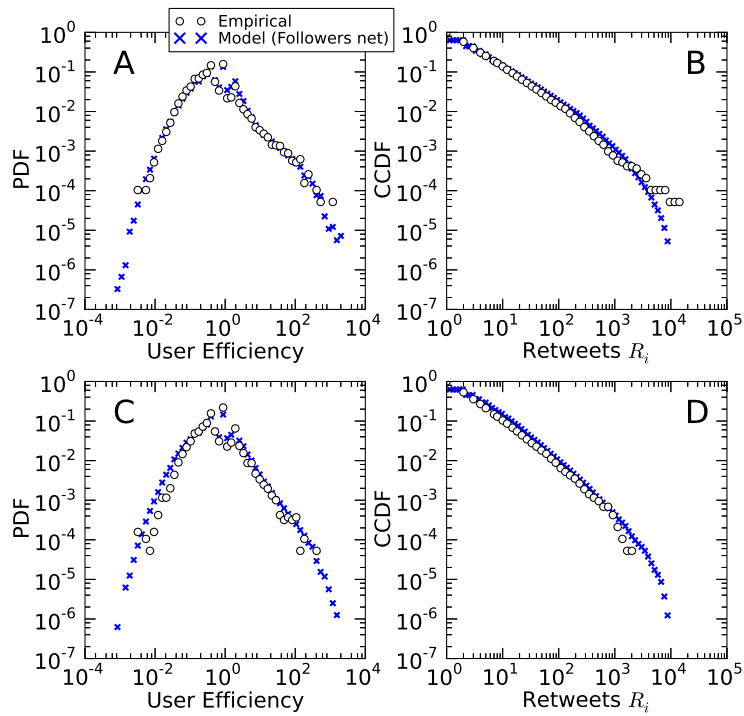}\\
 \caption{Model results to the user efficiency distribution (left column) and retweets gained by user distribution (right column), with the empirical results. The model has been applied to the followers network from the \#SOSInternetVE dataset (top panel) and the \#20N dataset (bottom panel).}
\label{Figure8a}
 \end{center}
\end{figure}

\begin{figure}
\begin{center}
 % replace aims_logo.eps by your figure file name
 \includegraphics[width=4in]{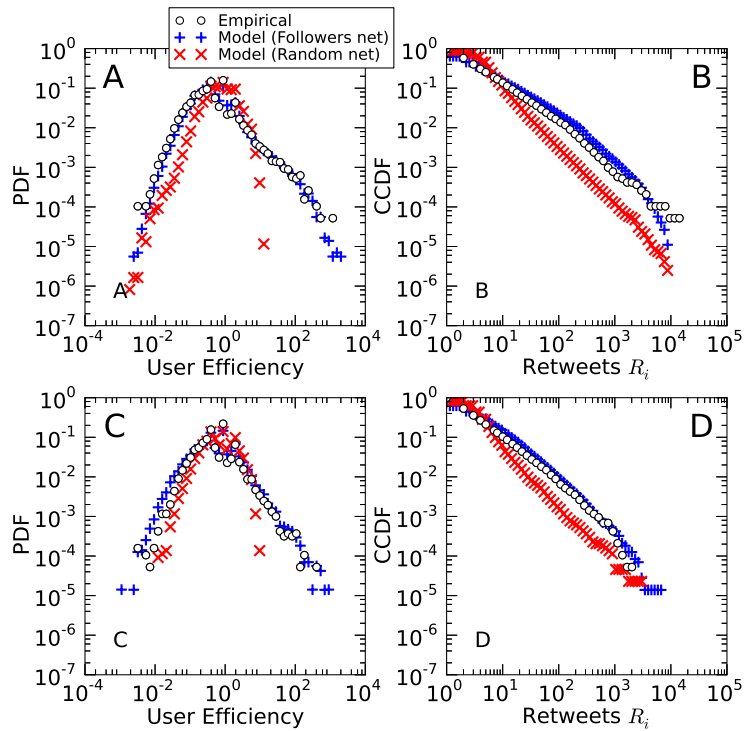}\\
 \caption{Effects of the underlying network topology on the model results in terms of the user efficiency distribution (left column) and retweets gained by user distribution (right column). The model has been applied to the followers network (blue crosses) and their randomized versions (red x symbols). Two datasets have been considered: \#SOSInternetVE (top panel) and \#20N (bottom panel). In all cases, an heterogeneous initial activity distribution $P(A_0) \propto A_0^{-1.4}$ has been considered.}
\label{Figure8b}
 \end{center}
\end{figure}

\begin{figure}
\begin{center}
 % replace aims_logo.eps by your figure file name
 \includegraphics[width=4in]{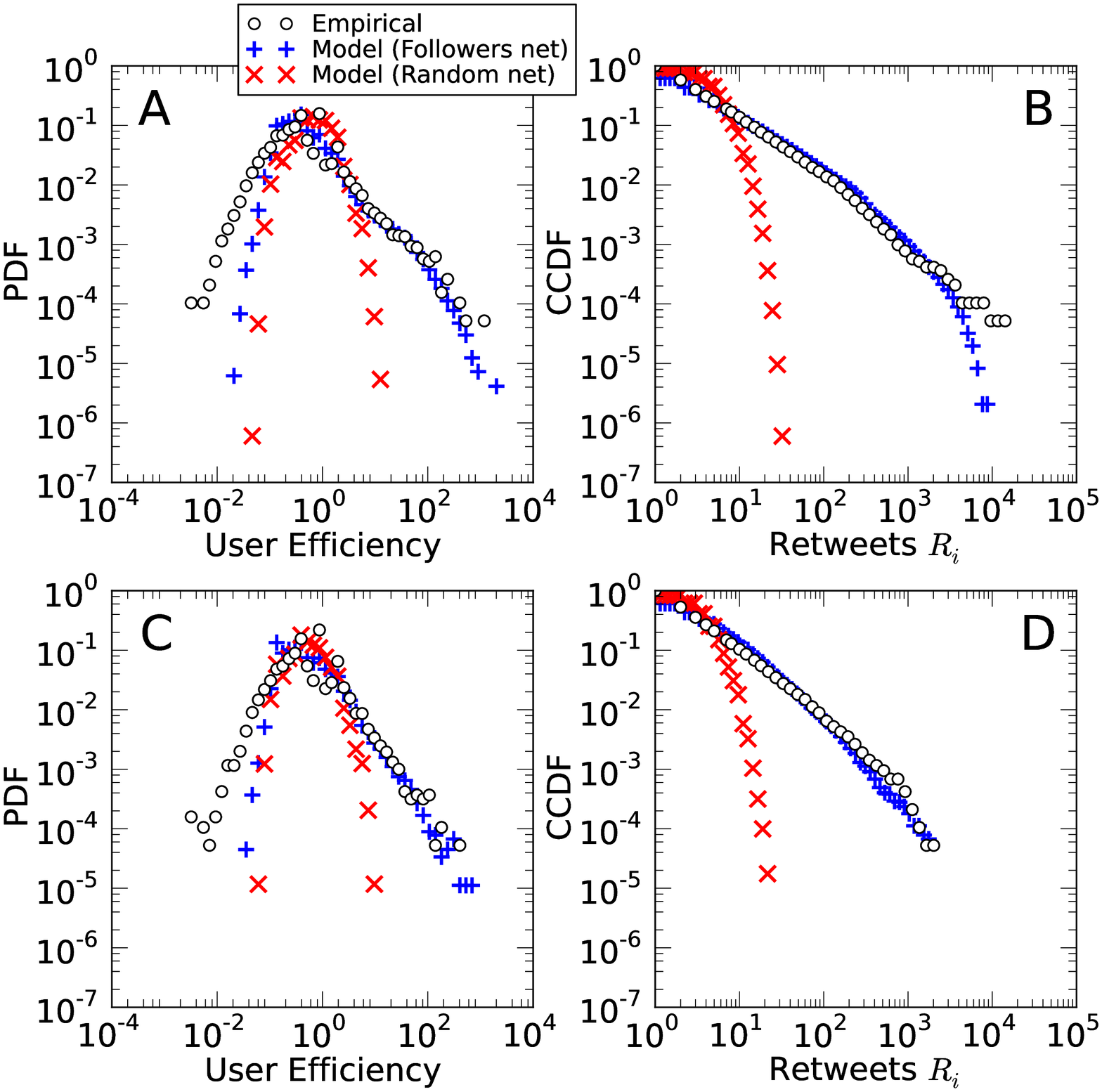}\\
 \caption{Effects of the individual user behavior on the model results in terms of the user efficiency distribution (left column) and retweets gained by user distribution (right column). The model has been applied to the followers network (blue crosses) and their randomized versions (red x symbols). Two datasets have been considered: \#SOSInternetVE (top panel) and \#20N (bottom panel). In all cases, an homogeneous activity distribution $P(A_0)=1/6$ where $A_0 \in [1,6]$ has been considered.}
\label{Figure8c}
 \end{center}
\end{figure}

{\color{black} We applied the model to two followers networks from the considered datasets. One of these networks corresponds to the present case study  {\it \#SOSInternetVE} and the other one is constructed from the {\it \#20N} dataset (see Fig. \ref{RiAall}D). The results of the user efficiency and retweets distribution are shown at the top and bottom panel in Fig. \ref{Figure8a} respectively. These results correspond to the average value of 50 model realizations. In both cases, the system has been initially excited using an heterogeneous user activity distribution in the form: $P(A_0) \propto A_0^{-1.4}$, and the spreading probabilities were taken from the cascade's characterization, given in Fig. \ref{CaracterizacionCadena}. It can be noticed that the resulting efficiency distributions in Fig. \ref{Figure8a}A and C (blue crosses) present a very good agreement with the empirical data (open circles) in both cases. In fact, the distributions also presents the different scaling behavior at the right side of the curve. Besides, the resulting retweets distributions in Fig. \ref{Figure8a}B and D (blue crosses), are also in very good agreement with the empirical data (open circles). These results show that the distributions analyzed are a reflection of the dynamical process behind the message spreading, which happens on Twitter by means of the retweets mechanism in independent cascades, where the probability of a cascade to grow decays as the message travels through the network, independently of the social context. After having validated the spreading mechanism, we are able to use the model to control the effect of the different factors that determine the user efficiency patterns, such as the heterogeneity of the underlying network topology and the characteristics of the individual user behavior (activity distribution).

First, we analyze the effects of the heterogeneity of underlying network topology on the spreading process. For this matter we applied the model to two different kind of substrata: the followers networks, from the datasets {\it \#SOSInternetVE} and {\it \#20N}, and their randomized versions. These randomized networks were built to avoid the presence of hubs and create homogeneous users profiles, by rewiring the edges so the degree distribution would follow a Normal curve instead of a power law, but maintaining the average number of edges per node. The resulting $\eta$ distributions after having excited the system with the same heterogeneous $P(A_0)$ are plotted by red x symbols in Fig. \ref{Figure8b}A andC respectively. It can be noticed that the distributions from these homogeneous networks present a different behavior than the ones obtained from the empirical observations and the modelled ones on the followers networks. There is a slightly lower density of the low efficient users, but more importantly, the highest values of the distribution are almost two orders below the empirical values, apparently following a lognormal behavior. However, the retweets distributions in Fig. \ref{Figure8b}B and D (red x symbols) still present power law behavior, due to the heterogeneity of  $P(A_0)$, although the probabilities of retweet are lower. In both cases, this means that an homogeneous society would allow users to gain an extremely high amount of retweets, only by means of employing an enormous amount of initial activity as well, since the user efficiency is strongly limited to the available connections on the underlying network.

Second, to study the effects of the individual user behavior, given by the initial activity distribution, we also applied the model to both followers networks (the case study {\it \#SOSInternetVE} and the {\it \#20N} dataset) and their randomized versions, but in this case considering an homogeneous $P(A_0)$, in the form: $P(A_0)=1/6$ where $A_0 \in [1,6]$, instead of the heterogeneous one previously considered. The results of applying this homogeneous user behavior to the heterogeneous followers networks are presented by blue crosses in Fig. \ref{Figure8c}. It can be noticed that the resulting user efficiency distributions in Fig. \ref{Figure8c}A and C, present the same behavior on the right side of the curve as the empirical observations (open circles), even though the considered user behavior is radically different than the empirical one. Besides, the retweets distribution (Fig. \ref{Figure8c}B and D)  also coincide quite well with the empirical observations and hardly changes in comparison to the distributions obtained when users posted messages in a heterogeneous way. However, if we change the substrata to their randomized versions, the model results no longer reproduce the empirical behavior and all the distributions loose their heterogeneity (red x symbols in Fig. \ref{Figure8c}). This confirms that the emerging patterns are not dependent on the way users post original messages, but instead a consequence of their heterogeneous connections on the underlying network.}

In the case of Twitter, the followers network also represents the way that the collective attention is organized. On this basis, this model has shown that if this collective attention is distributed heterogeneously among the population, the way users post messages has no further effects in the efficiency distribution, nor the retweets distribution, since the high aggregation of users around the influential ones is what produces such large collective reactions. In turn, if users would pay attention to each other homogeneously, as the randomized version of the followers network, then the retweets gained by user would be a reflection of the frequency and amount of posted messages, and the efficiency to gain such retweets would be strongly limited by the properties of the underlying substratum. However, despite the fact that in an homogeneous society it would be more difficult to find extreme cases of high efficient users, the density of extremely low efficient users also decreases when the attention is shared homogeneously among the collective. Therefore, this evidences that in order for some users to gain attention from the collective, others must loose it at the same time.
\\
\\
In summary, we have been able to model the efficiency of users to spread their opinions during Twitter conversations, and found that the emergent patterns are remarkable influenced by the underlying network topology. We have shown an evidence of the {\it robust but vulnerable} property of complex networks. In the sense that complex networks appear to be robust for most of the external excitations, as most of people post messages that do not travel at all, but vulnerable for selected excitations, as the activity performed by the highly efficient users have a remarkable impact in the resulting patterns \cite{Watts2002}. This effect is also  measured through the macroscopical property of the percentage of retweets on the overall posted messages. In the protest 47\% of the messages were retweets, while our simulations gave $45 \pm 3\%$ for the followers network and $40.3 \pm 0.1\%$ for the randomized version. This additional 5\% of retransmissions were only possible due to the complex organization of the network. 

%\begin{figure}
%\begin{center}
% % Requires \usepackage{graphicx}
% % replace aims_logo.eps by your figure file name
% \includegraphics[width=3.5in]{Figure8.eps}\\
% \caption{Results of the numerical simulations on the Randomized Network in terms of User Efficiency (A) and Retweets gained by user (B).}\label{Figure8}
% \end{center}
%\end{figure}

\section{Conclusions}

While spreading processes have been largely studied across several disciplines, accurate models to explain empirical dynamical processes are still an open field. In this paper we have performed a quantitative analysis of the structural and dynamical patterns of the activity on Twitter during an online political protest {\color{black} and generalized our results to other online conversations.} We found that the activity is fed by a small group of very active users, while the large majority hardly participated. As part of this activity there are interactions that determine the collective attention, which we found to be dominated by a very small group of highly influential users. However, if any, the rest of users gain influence in proportion to the activity they employ. Although, for the large majority of users the efforts are usually higher than the results. We propose a way to measure this bonding between actions and reactions, as the ratio between the retransmissions gained and user activity, that we understand as the individual efficiency to have messages spread in the network and hence it can be considered as a measure to be influential in the information spreading process. We found this measure to be universal across several Twitter conversations, as it is distributed following a lognormal distribution with a larger density of users at the higher orders, in all the studied cases. We propose a model to explain  {\color{black} the nature of }the efficiency distribution, based on biased independent cascades on the followers networks. The model results unveiled the effects of topology and individual behavior into the emergent dynamical patterns.  {\color{black} More particularly, it revealed that the emergence of a small fraction of highly efficient users results from the heterogeneity of the underlying network, rather than the differences in the individual user behavior. In fact, we found that in an homogeneously organized society we would need a much larger population to find the same level of influence to diffuse information that we get by complex and heterogeneous organizing.} We conclude that although individuals may have remarkable psychological and contextual differences, the dynamical patterns are due to simple and universal interaction mechanisms.

 \section*{Acknowledgement}
This research was supported by the Ministry of Economy and Competitiveness-Spain under Grant No. MTM2012-39101


\begin{thebibliography}{10}

\bibitem{TelecomEconomist}
Research Infiniti.
\newblock Global revenue assurance market in the telecom sector 2012-2016.
\newblock White Paper, 2013.

\bibitem{BigDataThirdWorld}
Systems Cisco.
\newblock Cisco visual networking index: Forecast and methodology, 2012–2017.
\newblock white Paper, 2013.

\bibitem{Lazer2009}
David Lazer, Alex Pentland, Lada Adamic, Sinan Aral, Albert-Laszlo Barabasi,
  Devon Brewer, Nicholas Christakis, Noshir Contractor, James Fowler, Myron
  Gutmann, Tony Jebara, Gary King, Michael Macy, Deb Roy, and Marshall~Van
  Alstyne.
\newblock Social science: Computational social science.
\newblock {\em Science}, 323(5915):721--723, February 2009.

\bibitem{Lewis2008}
Kaufman J. Gonzalez M. Wimmer~A. Lewis, K. and N.~Christakis.
\newblock {Tastes, ties, and time: A new social network dataset using
  Facebook.com}.
\newblock {\em Social Networks}, 30(4):330--342, October 2008.

\bibitem{Takhteyev2012}
Yuri Takhteyev, Anatoliy Gruzd, and Barry Wellman.
\newblock {Geography of Twitter networks}.
\newblock {\em Social Networks}, 34(1):73--81, January 2012.

\bibitem{Mislove2011}
Alan Mislove, Sune Lehmann, Yong-Yeol Ahn, Jukka-Pekka Onnela, and J.~Niels
  Rosenquist.
\newblock Understanding the demographics of twitter users.
\newblock In Lada~A. Adamic, Ricardo~A. Baeza-Yates, and Scott Counts, editors,
  {\em ICWSM}. The AAAI Press, 2011.

\bibitem{Gayo2012}
Daniel Gayo-Avello.
\newblock "i wanted to predict elections.
\newblock {\em CoRR}, abs/1204.6441, 2012.

\bibitem{Zeng2011}
Johan Bollen, Huina Mao, and Xiao-Jun Zeng.
\newblock Twitter mood predicts the stock market.
\newblock {\em J. Comput. Science}, 2(1):1--8, 2011.

\bibitem{Matsuo2011}
Takeshi Sakaki, Makoto Okazaki, and Yutaka Matsuo.
\newblock Earthquake shakes twitter users: real-time event detection by social
  sensors.
\newblock In {\em Proceedings of the 19th international conference on World
  wide web}, WWW '10, pages 851--860, New York, NY, USA, 2010. ACM.

\bibitem{Culotta2010}
Aron Culotta.
\newblock Towards detecting influenza epidemics by analyzing twitter messages.
\newblock In {\em Proceedings of the First Workshop on Social Media Analytics},
  SOMA '10, pages 115--122, New York, NY, USA, 2010. ACM.

\bibitem{Borondo2012}
J. Borondo, A. J. Morales, J. C. Losada, R. M. Benito,
\newblock Characterizing and modeling an electoral campaign in the context of Twitter: 2011 Spanish Presidential election as a case study
\newblock  {\em Chaos}, 22 (2), 023138, doi 10.1063/1.4729139,  2012


\bibitem{Livne2011}
Avishay Livne, Matthew~P. Simmons, Eytan Adar, and Lada~A. Adamic.
\newblock The party is over here: Structure and content in the 2010 election.
\newblock In Lada~A. Adamic, Ricardo~A. Baeza-Yates, and Scott Counts, editors,
  {\em ICWSM}. The AAAI Press, 2011.

\bibitem{Sysomos}
Alex Cheng and Mark Evans.
\newblock An in-depth look inside the twitter world.
\newblock WWW page, 2009.

\bibitem{Wang2011}
Sitaram Asur, Bernardo~A. Huberman, Gábor Szabó, and Chunyan Wang.
\newblock Trends in social media: Persistence and decay.
\newblock {\em CoRR}, abs/1102.1402, 2011.

\bibitem{Romero:2010tj}
Daniel~M. Romero, Wojciech Galuba, Sitaram Asur, and Bernardo~A. Huberman.
\newblock Influence and passivity in social media.
\newblock In {\em Proceedings of the ECML/PKDD 2011}, 2011.

\bibitem{Watts2011}
Eytan Bakshy, Jake~M. Hofman, Winter~A. Mason, and Duncan~J. Watts.
\newblock Everyone's an influencer: quantifying influence on twitter.
\newblock In {\em Proceedings of the fourth ACM international conference on Web
  search and data mining}, WSDM '11, pages 65--74, New York, NY, USA, 2011.
  ACM.

\bibitem{Goldenberg2001}
Jacob Goldenberg, Barak Libai, and Eitan Muller.
\newblock Talk of the network: A complex systems look at the underlying process
  of word-of-mouth.
\newblock {\em Marketing Letters}, 2001.

\bibitem{Kwak2010}
Haewoon Kwak, Changhyun Lee, Hosung Park, and Sue Moon.
\newblock What is twitter, a social network or a news media?
\newblock In {\em WWW '10: Proceedings of the 19th international conference on
  World wide web}, pages 591--600, New York, NY, USA, 2010. ACM.

\bibitem{Huberman2009}
Bernardo~A. Huberman, Daniel~M. Romero, and Fang Wu.
\newblock Social networks that matter: Twitter under the microscope.
\newblock {\em First Monday}, 14(1), 2009.

\bibitem{Lotan2010}
Danah Boyd, Scott Golder, and Gilad Lotan.
\newblock Tweet, tweet, retweet: Conversational aspects of retweeting on
  twitter.
\newblock In {\em HICSS}, pages 1--10. IEEE Computer Society, 2010.

\bibitem{Despotovic2010}
Wojciech Galuba, Karl Aberer, Dipanjan Chakraborty, Zoran Despotovic, and
  Wolfgang Kellerer.
\newblock Outtweeting the twitterers - predicting information cascades in
  microblogs.
\newblock In {\em Proceedings of the 3rd conference on Online social networks},
  WOSN'10, pages 3--3, Berkeley, CA, USA, 2010. USENIX Association.

\bibitem{Cha2010}
Meeyoung Cha, Hamed Haddadi, Fabrício Benevenuto, and Krishna~P. Gummadi.
\newblock Measuring user influence in twitter: The million follower fallacy.
\newblock In {\em in ICWSM ’10: Proceedings of international AAAI Conference
  on Weblogs and Social}, 2010.

\bibitem{Xiong2012}
Fei Xiong, Yun Liu, Zhen jiang Zhang, Jiang Zhu, and Ying Zhang.
\newblock An information diffusion model based on retweeting mechanism for
  online social media.
\newblock {\em Physics Letters A}, 376(30–31):2103 -- 2108, 2012.

\bibitem{RomeroHT}
Daniel~M. Romero, Chenhao Tan, and Johan Ugander.
\newblock {Social-Topical Affiliations: The Interplay between Structure and
  Popularity}, December 2011.

\bibitem{LehmannHT}
Janette Lehmann, Bruno Gon\c{c}alves, Jos{\'e}~J. Ramasco, and Ciro Cattuto.
\newblock Dynamical classes of collective attention in twitter.
\newblock In {\em Proceedings of the 21st international conference on World
  Wide Web}, WWW '12, pages 251--260, New York, NY, USA, 2012. ACM.

\bibitem{Morales2012}
A. J. Morales, J. C. Losada, R. M. Benito,
\newblock  Users structure and behavior on an online social network during a political protest
\newblock {\em Physica A: Statistical Mechanics and its Applications}, 391 (21) 5244-5253 (2012), doi 10.1016/j.physa.2012.05.015, 2012.

\bibitem{Newman2006}
M.~E.~J. Newman.
\newblock {Power laws, Pareto distributions and Zipf's law}.
\newblock {\em Contemporary Physics}, 46(5):323--351, May 2005.

\bibitem{Newman2003}
M.~E.~J. Newman.
\newblock Mixing patterns in networks.
\newblock {\em Physical Review E}, 67(2):026126, 2003.

\bibitem{Hu2009}
Hai-Bo Hu and Xiao-Fan Wang.
\newblock Disassortative mixing in online social networks.
\newblock {\em EPL (Europhysics Letters)}, 86(1):18003, 2009.

\bibitem{Barabasi2011}
Dashun Wang, Zhen Wen, Hanghang Tong, Ching-Yung Lin, Chaoming Song, and
  Albert-Laszlo Barabasi.
\newblock Information spreading in context.
\newblock In {\em Proceedings of the 20th international conference on World
  wide web}, WWW '11, pages 735--744, New York, NY, USA, 2011. ACM.

\bibitem{Wu2004}
F.~Wu, B.A. Huberman, L.A. Adamic, and J.R. Tyler.
\newblock {Information flow in social groups}.
\newblock {\em Physica A: Statistical Mechanics and its Applications},
  337(1-2):327--335, 2004.

\bibitem{Mitzenmacher2004}
M.~Mitzenmacher.
\newblock A brief history of generative models for power law and lognormal
  distributions.
\newblock {\em Internet Mathematics}, 1(2):226--251, 2004.

\bibitem{Moro2011}
José~Luis Iribarren and Esteban Moro.
\newblock Affinity paths and information diffusion in social networks.
\newblock {\em Social Networks}, 33(2):134 -- 142, 2011.

\bibitem{Moro2011b}
Jose~Luis Iribarren and Esteban Moro.
\newblock Branching dynamics of viral information spreading.
\newblock {\em Phys. Rev. E}, 84:046116, 2011.

\bibitem{Watts2002}
Duncan~J. Watts.
\newblock {A simple model of global cascades on random networks}.
\newblock {\em Proceedings of the National Academy of Sciences},
  99(9):5766--5771, April 2002.

\end{thebibliography}
\end{document}